\documentclass[sigconf]{acmart}

\acmConference[ICSE 2024]{46th International Conference on Software Engineering}{April 2024}{Lisbon, Portugal}
\usepackage{soul}
\usepackage[edges]{forest}
\usepackage{forest}
\usepackage{amsmath}
\usepackage{svg}
\usepackage{algorithmic}
\usepackage{textcomp}
\usepackage{xcolor}
\usepackage{subfiles}
\usepackage{wrapfig}
\usepackage[utf8]{inputenc}
\usepackage{graphicx}
\usepackage{url}
\usepackage{enumitem}
\usepackage{balance}
\usepackage{booktabs}
\usepackage{multicol}
\usepackage{multirow}
\usepackage{tabularx}
\usepackage{longtable}
\usepackage{array}
\usepackage{threeparttable}
\usepackage{multirow}
\usepackage{subcaption}
\usepackage{amsthm}
\usepackage{pifont}
\usepackage{xspace}
\usepackage{tikz}
\usetikzlibrary{trees}
\usepackage{tcolorbox}
\usepackage{geometry}
\usepackage{framed,color,verbatim}
\usepackage{array}
\usepackage{tikz}
\usepackage[T1]{fontenc}
\usepackage[utf8]{inputenc}
\usepackage{hyperref}
\usepackage{colortbl}
\definecolor{shadecolor}{rgb}{.9, .9, .9}

\graphicspath{ {./figs/} }

\newboolean{showcomments}
\setboolean{showcomments}{true}
\ifthenelse{\boolean{showcomments}}
 { \newcommand{\mynote}[2]{
      \fbox{\bfseries\sffamily\scriptsize#1}
        {\small$\blacktriangleright$\textsf{\emph{#2}}$\blacktriangleleft$}}}
        { \newcommand{\mynote}[2]{}}

\newcommand{\sectopic}[1]{\vspace{0.2em}\par\noindent{\textit{\bfseries #1}}}

\def\BibTeX{{\rm B\kern-.05em{\sc i\kern-.025em b}\kern-.08em
    T\kern-.1667em\lower.7ex\hbox{E}\kern-.125emX}}

\copyrightyear{2024}
\acmYear{2024}
\setcopyright{rightsretained}
\acmConference[ICSE-SEIP '24]{46th International Conference on Software Engineering: Software Engineering in Practice}{April 14--20, 2024}{Lisbon, Portugal}
\acmBooktitle{46th International Conference on Software Engineering: Software Engineering in Practice (ICSE-SEIP '24), April 14--20, 2024, Lisbon, Portugal}\acmDOI{10.1145/3639477.3639732}
\acmISBN{979-8-4007-0501-4/24/04}

\title{Enhancing Text-to-SQL Translation for Financial System Design}

\author{Yewei Song}
\affiliation{
  \institution{University of Luxembourg}
  \country{Luxembourg}
}\email{yewei.song@uni.lu}
 
\author{Saad Ezzini}
\affiliation{
  \institution{Lancaster University}
  \country{United Kingdom}
}\email{s.ezzini@lancaster.ac.uk}

\author{Xunzhu Tang}
\affiliation{
  \institution{University of Luxembourg}
  \country{Luxembourg}
}\email{xunzhu.tang@uni.lu}
\author{Cedric Lothritz}
\affiliation{
  \institution{University of Luxembourg}
  \country{Luxembourg}
}\email{cedric.lothritz@uni.lu}
\author{Jacques Klein}
\affiliation{
  \institution{University of Luxembourg}
  \country{Luxembourg}
}\email{jacques.klein@uni.lu}

\author{Tegawendé Bissyandé}
\affiliation{
  \institution{University of Luxembourg}
  \country{Luxembourg}
}\email{tegawende.bissyande@uni.lu}

\author{Andrey Boytsov}
\affiliation{
  \institution{Banque BGL BNP Paribas}
  \country{Luxembourg}
}\email{andrey.boytsov@bgl.lu}
\author{Ulrick Ble}
\affiliation{
  \institution{Banque BGL BNP Paribas}
  \country{Luxembourg}
}\email{ulrick.ble@bgl.lu}
\author{Anne Goujon}
\affiliation{
  \institution{Banque BGL BNP Paribas}
  \country{Luxembourg}
}\email{anne.goujon@bgl.lu}

\begin{document}
\begin{abstract}


Text-to-SQL, the task of translating natural language questions into SQL queries, is part of various business processes. Its automation, which is an emerging challenge, will empower software practitioners to seamlessly interact with relational databases using natural language, thereby bridging the gap between business needs and software capabilities.

In this paper, we consider Large Language Models (LLMs), which have achieved state of the art for various NLP tasks. Specifically, we benchmark Text-to-SQL performance, the evaluation methodologies, as well as input optimization (e.g., prompting). In light of the empirical observations that we have made, we propose two novel metrics that were designed to adequately measure the similarity between SQL queries. 

Overall, we share with the community various findings, notably on  how to select the right LLM on Text-to-SQL tasks. We further demonstrate that a tree-based edit distance constitutes a reliable metric for assessing the similarity between generated SQL queries and the oracle for benchmarking Text2SQL approaches. This metric is important as it relieves researchers from the need to perform computationally expensive experiments such as executing generated queries as done in prior works. Our work implements financial domain use cases and, therefore contributes to the advancement of Text2SQL systems and their practical adoption in this domain.

\end{abstract}

\settopmatter{printacmref=false}

\maketitle

\section{Introduction} \label{sec:introduction}
Advances in natural language processing (NLP), notably with the advent of large language models (LLMs), have led to significant performance improvements in various Text-to-Code generation tasks. 
Among these, Text-to-SQL, i.e., the process of translating natural language queries into SQL queries, is an emerging task, with various applications in businesses. Implementing reliable Test-to-SQL automation is expected to empower practitionners, i.e., users and non-tech agents, to seamlessly interact with business relational databases using their own natural language.  Breaking through the challenge of translating text to SQL code will bridge an important gap between business needs and software capabilities, and unlock new possibilities for software integration.

Our work focuses on Text2SQL with a specific emphasis on the financial domain.  Leveraging the promising capabilities of LLMs, which have garnered widespread adoption across various domains, including code generation, our research endeavors to showcase the tackle several pressing challenges in this field.

To this end, this paper contributes to the advancement of Text2SQL systems and their practical utility. We turn our attention to three primary areas: performance benchmarking, evaluation methodologies, and input optimization. Within these domains, we propose two novel metrics designed to measure SQL query similarity, demonstrating impressive efficacy. Additionally, we conduct rigorous benchmarking exercises to evaluate the performance of LLMs while also exploring the impact of rephrased questions and two distinct prompt types. Furthermore, we curate a specialized sub-dataset focused on bank operations, replete with complex questions that pose a formidable challenge to state-of-the-art models. In doing so, our work not only sheds light on the potential enhancements to Text2SQL systems but also offers valuable insights into the broader landscape of NLP applications in software engineering, especially in domains with complex and domain-specific requirements.

\sectopic{Contributions.} 
In this paper, we present three primary contributions. Firstly, we perform an empirical study on a publicly available dataset for the Text2SQL task and benchmark several SOTA LLMs. We provide some insights into model selection in the financial domain. Secondly, we create a set of new and challenging test cases specifically tailored to transaction-related queries, filling a critical gap in the availability of realistic dataset resources for banking operations. Thirdly, we propose two innovative evaluation metrics designed to accurately assess the performance of Text2SQL models. These metric exhibits a high correlation with execution matching, eliminating the need to execute code on the database and offering more efficient and practical means of model evaluation in the context of banking applications. Additionally, we propose question optimization methods that improve performance for Text2SQL models.


\sectopic{Findings.} Our work holds 2 important findings for practice: First, we conclude that for Text-to-SQL applications in financial business scenarios, choosing LLM with more code training data has better performance, such as \textit{nsql-6B} and \textit{CodeGen2}. Second, we found that the new metric we proposed, Tree Similarity of Editing Distance(TSED), is currently the best Text-to-SQL evaluation metric when the original database cannot be used for result evaluation.


\section{Background and Related Work} \label{sec:background}
This section presents the necessary background for our solutions and further discusses the related literature in SE and NLP.  

\sectopic{Large Language Models}.
LLMs, including models like GPT-4 and PaLM, represent a transformative advancement in the field of NLP. These models are characterized by their extensive scale, boasting millions to billions of parameters, enabling them to learn intricate linguistic patterns and relationships from vast amounts of text data. This capacity allows them to generate human-like text and perform a wide array of language-related tasks, from text generation and machine translation to sentiment analysis and question answering. Their pre-trained nature, combined with fine-tuning on specific tasks, has led to groundbreaking progress in NLP. As LLMs continue to evolve, they promise to redefine the landscape of NLP by fostering more nuanced understanding, interaction, and generation of text. 

\begin{figure*}
    \centering
    \includegraphics[width=11cm]{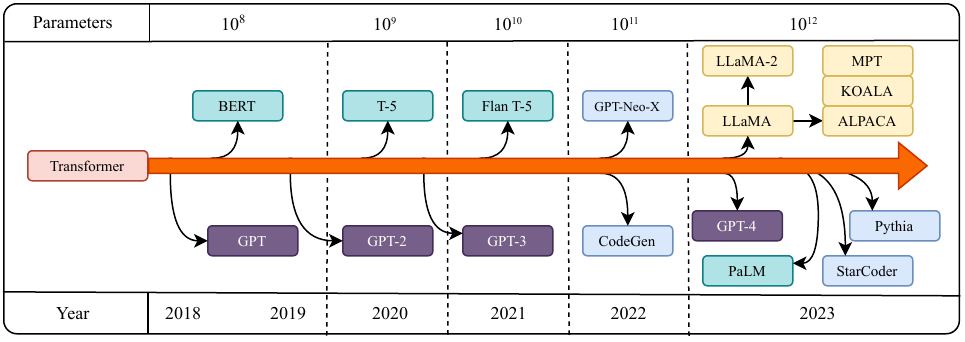}
    \vspace{-3mm}
    \caption{Development of Large Language Models from Year 2018-2023}
    \label{fig:Phylogenetic}
    \vspace{-4mm}
\end{figure*}

The development of LLMs finds a common ancestry in Transformer proposed in 2017~\cite{vaswani2017attention}, as illustrated in Figure \ref{fig:Phylogenetic}.
These models have diversified due to differences in deep neural network structure and training strategies, with diverse optimization and supervised fine-tuning techniques like LoRA~~\cite{hu2021lora} and P-Tuning v2~~\cite{liu2022ptuning} alleviating training complexities. The HuggingFace LLM leaderboard~\cite{huggingfaceLeaderBoard}, for instance, featured approximately 75 models in May 2023, but by mid-August 2023, this number had surged to over 600, marking the "explosion of LLMs" in the field. A recurring trend in contemporary LLMs is the exponential growth in the number of model parameters, with models like GPT-4 (presumed) expanding their parameter count by up to $10^5$ times when compared to the BERT model, as depicted in Figure \ref{fig:Phylogenetic}. The question of whether this increase in model size translates to diminishing marginal returns in terms of performance remains an enigma, but some papers, such as LLaMA~~\cite{touvron2023llama} and Chinchilla~~\cite{hoffmann2022training}, have suggested that the pre-training corpus may need to scale proportionally to the number of parameters, a practice not always followed by modern LLMs.

\sectopic{The Text-to-SQL task,} serving as a vital link between human language and structured databases, has gained prominence in the field of natural language processing. Some survey papers, including contributions by Qin~\cite{qin2022survey} and Katsogiannis~\cite{katsogiannis2023survey}, introduce some of the latest advances up to the first half of 2023. It involves translating natural language queries into structured SQL queries for database execution, aligning with the imperative for AI systems to understand and process user queries in a human-like manner. This task has its roots in semantic parsing and question answering, initially explored through rule-based and template-based approaches. However, the advent of deep learning, especially sequence-to-sequence models like Seq2Seq with attention mechanisms, marked a significant turning point, enabling more precise and adaptable conversion of natural language into SQL. Datasets such as WikiSQL and ATIS established standardized benchmarks for evaluating Text2SQL models~\cite{zhong2017seq2sql,hemphill1990atis}, while the SPIDER dataset~\cite{yu2018spider}, meticulously annotated by Yale students, further accelerated progress.

In the Text2SQL domain, cutting-edge models have emerged to address the challenge of converting natural language queries into SQL queries. PICARD, introduced by Scholak et al.~\cite{scholak2021picard}, combines a beam-search-like processor with the T5 model to effectively understand and generate SQL queries. Some remarkable post-PICARD advancements in existing methodologies, exemplified by RESDSQL~\cite{li2023resdsql} and Graphix-3B~\cite{li2023graphix}. However, it remains essential to scrutinize the performance of LLMs, particularly in Zero-Shot settings. Anticipating ongoing optimization and fine-tuning on Text2SQL datasets, it is evident that this task continues to evolve and hold a significant position in the NLP landscape.

In the realm of natural language processing, several Text2SQL datasets have emerged as pivotal resources for advancing the capabilities of semantic parsing models. Datasets such as WikiSQL, ATIS, and the recently introduced CoSQL, each contribute unique challenges and linguistic nuances from various domains~\cite{zhong2017seq2sql,dahl1994expanding,yu2019cosql}. However, the SPIDER dataset stands out due to its expansive complexity and cross-domain nature, making it a preferred choice for research and development in this field~\cite{yu2018spider}. Notably, the SPIDER dataset benefits from the involvement of 7 Yale students who meticulously annotated the data, enhancing its accuracy and quality.

In Text2SQL evaluation, two crucial metrics are employed from SPIDER paper~\cite{yu2018spider}: Execution Match, which assesses query utility by executing it against a database; Exact Match, which measures both semantic and syntactic accuracy. Besides, the BLEU score, adapted from machine translation evaluation, gauges query fluency and relevance~\cite{papineni-etal-2002-bleu}. Additionally, the use of Abstract Syntax Trees (AST) aids in evaluating structural similarity~\cite{koschke2006clone}, offering deeper insights into a system's ability to capture underlying query structures.

\vspace{-2mm}
\section{Scenario and Challenges}~\label{sec:approach}
Together with our industrial partner from the financial domain, we aimed to enhance a question-answering chatbot. To that end, we have incorporated a Text2SQL module. This module serves as a crucial component, enabling smooth interactions and ensuring our chatbot's ability to accurately retrieve data from the relational database and construct precise responses. The Text2SQL module acts as an intermediary, converting natural language queries from users into SQL queries that can be directly executed on the bank's database. Then we need to add an evaluation part to the module to check if the results are correct or not. This allows the company can easily integrate it into its chatbot system and keep polishing it.

\vspace{-1mm}
\subsection{Text to SQL Pipeline}
\vspace{-1mm}

\begin{figure}[H]
\centering
\includegraphics[width=\columnwidth]{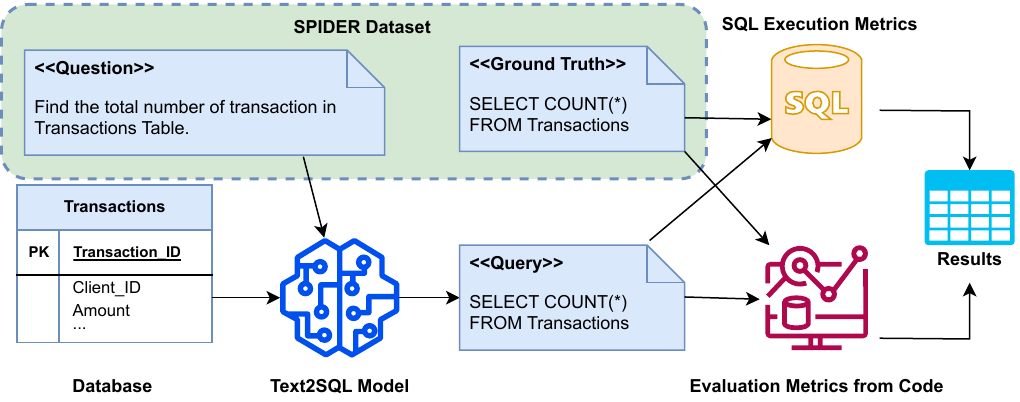}
\vspace{-3mm}
\caption{Pipeline of Text to SQL Evaluation Practice}
\label{fig:pipeline}
\end{figure}

This paper aims to assess the performance of several LLMs on the Text2SQL task by considering practical constraints. These constraints lead to several challenges that are elicited in Section~\ref{subsec:challenges}.
In this section, we detail the Text2SQL pipeline used to assess the various LLMS. 
The entire pipeline is shown in Figure \ref{fig:pipeline}. 
For several of our experiments, we rely on the SPIDER dataset~\cite{yu2018spider}, which provides databases, questions in natural language, and the corresponding "ground truth" SQL queries. 
For a given question, the goal of the Text2SQL model is to generate a SQL query that generates the same result as the SQL query from the Ground Truth. 
To assess the quality of the generated SQL queries, several metrics are computed (detailed in Section \ref{subsec:metrics}). Some of these metrics are computed directly from the code (i.e., we compare the generated SQL queries against the ground truth queries), and others are computed based on the result of the execution of the queries. 
The pipeline also includes some key steps inside the Text2SQL model:
\begin{itemize}
    \item \textbf{Preprocessing}: The incoming natural language query is tokenized and transformed into numerical representations as embedding vectors.
    \item \textbf{Encoding}: The tokenized and numerical representation of the natural language query is fed into the trained Encoder of the Seq-to-Seq model.
    \item \textbf{Context Vector}: The Encoder processes the input query and generates a fixed-size context vector that captures the semantic information of the query.
    \item \textbf{Decoding}: The context vector is passed to the trained Decoder of the Seq-to-Seq model. The Decoder generates the SQL query token by token based on the information encoded in the context vector.
    \item \textbf{SQL Query}: As the Decoder generates SQL tokens, they are combined to form the final SQL query.
\end{itemize}
The Text2SQL module, using a Seq-to-Seq model, directly learns the mapping between natural language queries and SQL queries. Through a process of encoding the natural language input and decoding the corresponding SQL output, it enables seamless communication between users and the bank's database, allowing users to interact using natural language queries via the chatbot interface.

\subsection{Challenges}\label{subsec:challenges}
The implementation of a Text2SQL module for a banking chatbot comes with several challenges. 
In this subsection, we discuss the main challenges and potential approaches to address them:

\sectopic{Model Selection:} With the emergence of new Language Model architectures, such as LLMs mentioned in Section 2, it becomes challenging to determine which model would yield the best performance for the Text2SQL task. Selecting the most appropriate model requires careful consideration of factors like model size, training data, computational resources, and performance on the specific task. Conducting comparative experiments and model evaluations will be essential to identify the optimal model for our use case.

\vspace{-2mm}
\vspace{-3mm}

\sectopic{Optimizing Generated Questions:} The chatbot system's generated questions may not always be in the most effective format for the Text2SQL module. Optimizing the generated questions to align better with the model's input expectations can improve overall performance. Techniques such as data augmentation, paraphrasing, or using intermediate templates may be explored to enhance the quality of user queries and ensure better Text2SQL conversion.

\sectopic{Cost-Performance Trade-off:} Most of the companies and research groups have limited computing and financial resources. Thus, achieving a high performance-to-cost ratio is crucial. LLMs and extensive training datasets can be computationally expensive and resource-intensive. Exploring smaller, more efficient models, and leveraging transfer learning techniques could be potential solutions to balance performance and resource constraints. In addition to cost considerations, solutions that consume fewer resources also offer environmental advantages.

\sectopic{Evaluation Metrics and Database Access:} Evaluating the performance of the Text2SQL module becomes challenging when direct access to the real database is limited or restricted, like when we working on a bank database. It is essential to define suitable evaluation metrics that can measure the module's effectiveness without the need for executing queries on the actual database. Metrics such as logical form matching and semantic correctness can provide insights into the module's performance even when direct execution is not feasible.

\sectopic{Limited Banking Domain Dataset:} Availability of a sufficient dataset specifically tailored to the banking domain is crucial for training a reliable Text2SQL model. Collecting domain-specific data can be time-consuming and may require domain experts. Leveraging transfer learning with pre-trained models on larger datasets and incorporating domain-specific fine-tuning can help mitigate the data scarcity issue.\\
To address these challenges effectively, we propose a combination of strategies, including thorough experimentation with different models, techniques for question optimization, efficient resource utilization, well-defined evaluation metrics, and creative data augmentation methods. Additionally, collaborating with domain experts and exploring partnerships with other organizations in the banking domain could help access more relevant data and resources to improve the Text2SQL module's performance for the bank's chatbot.

\newenvironment{code}%
   {\snugshade\verbatim}%
   {\endverbatim\endsnugshade}

\section{Empirical Set Up} \label{sec:evaluation}

In this section, we propose the research questions and our practice set up to resolve them. 

\subsection{Research Questions} \label{subsec:RQs}
Our evaluation tackles the following 5 research questions (RQs):

\sectopic{RQ1. } Which LLM is the most accurate for the Text2SQL task?
\sectopic{RQ2. } What is the effect of the question rephrasing, prompt optimization, and post-processing steps on the performance of LLMs?
\sectopic{RQ3. } Which LLM is the most efficient regarding execution time and computational resources?
\sectopic{RQ4. } When the database is not reachable, can we propose metrics to evaluate the quality of the generated SQL queries that are as accurate as the Execution Match metric?
\sectopic{RQ5. } Text2SQL tasks in financial scenarios (especially banks) are lacking in existing datasets. How should we solve it?

\subsection{Model Selection}

Our selection of LLMs is basically based on the selection of metrics in three directions, including the overall ranking list of LLMs, the ranking of Spider datasets, and LLMs with superior performance on Stack datasets.

As the undisputed performance ceiling, we choose ChatGPT based on GPT-4 as one of our research targets, and an interesting topic is how big the gap between other models and ChatGPT and GPT-4~\cite{bubeck2023sparks} or GPT-3.5-Turbo.
\sectopic{HuggingFace Ranking}
After LLMs gained enough attention, the HuggingFace platform launched a LeaderBoard\footnote{https://huggingface.co/spaces/HuggingFaceH4/open\_llm\_leaderboard}, where we observed that some models are considered 'breakthroughs' in this ranking, while others are optimized and fine-tuned versions of these models~\cite{open-llm-leaderboard}. In order to avoid interference, we chose the original models (standalone releases) as we can find it. We also selected the most current models according to Figure \ref{fig:Phylogenetic} of the LLM development history, including LLaMA~\cite{touvron2023llama}, GPT-NeoX, and so on. Based on the background above, we have chosen six LLMs to test against OpenAI models: MPT~\cite{MosaicML2023Introducing}, ALPACA~\cite{alpaca}, KOALA~\cite{koala_blogpost_2023}, OpenAssistant-Pythia~\cite{kopf2023openassistant}~\cite{biderman2023pythia}, LLaMA-2~\cite{touvron2023llama2}, and ORCA~\cite{mukherjee2023orca}.
\sectopic{Spider Dataset Leaderboard}
Our test SPIDER dataset has a frequently updated benchmark performance ranking\footnote{https://yale-lily.github.io/spider}. We selected the two models PICARD~\cite{scholak2021picard} and RESDSQL~\cite{li2023resdsql} as a baseline of non-LLM models.
\sectopic{The Stack Database}
The Stack dataset is one of the richest code-based corpora containing programming languages available. Therefore, we selected several LLMs as candidates based on their performance on the Stack dataset~\cite{Kocetkov2022TheStack}. We used several typical models as the Supervised group (models that have been pre-trained on programming languages), taking into account the existing research like CodeGen2~\cite{nijkamp2023codegen2}, StarCoder~\cite{li2023starcoder}, and nsql~\cite{numbersstation2023NSText2SQL}.

\subsection{Data Selection}
We run our experiments on two Datasets.

 \textbf{\ding{182} The SPIDER Dataset:}
We selected the "\texttt{dev}" evaluation subset from the SPIDER dataset for our evaluation. This subset contains 20 databases, with questions in natural language together with the corresponding ground truth SQL queries. The 20 databases cover four domains: Business (4), Entertainment (7), Education (4), and Travel and Leisure (5). Ground-truth queries in this subset are executable, as confirmed in related work. The databases are evenly distributed, with nearly half (9) having four tables, 4 containing three tables or fewer, 4 featuring five tables, and the remaining databases consisting of five or more tables. This diverse dataset allows for a thorough evaluation of Text2SQL models across different domains and database complexities.

\textit{Questions in natural language and Difficulty Level:}
The Text2SQL task manifests varying degrees of complexity depending on the difficulty of the questions in natural language.  
The questions are spread over four distinct difficulty levels:

\noindent
 \textbf{Level 1:} This simplest level involves uncomplicated queries where users retrieve straightforward information from a single table.

\noindent
 \textbf{Level 2:} Moving up the complexity ladder, Level 2 introduces basic statistical operations such as SUM, AVG, and COUNT. This requires the translation of these arithmetic operations into corresponding SQL expressions.

\noindent
 \textbf{Level 3:} Progressing to Level 3, the complexity deepens with the inclusion of GROUP BY clauses. This necessitates an understanding of aggregate functions applied to grouped data.

\noindent
 \textbf{Level 4:} The highest echelon of complexity is reached at Level 4, where JOIN operations are introduced. This challenges the model to grasp relationships between multiple tables and generate intricate SQL statements that seamlessly connect data from different sources.

After classification, we have 267 level 1 questions, 239 level 2 questions, 120 level 3 questions, 408 level 4 questions in SPIDER dev-set.

\textbf{\ding{183} A new small size Financial Dataset:}
We designed a  specific dataset for the Text2SQL task on the financial scene, which is a bank transactions database with 30 questions. 
We aim to challenge all SOTA models and find the gap between in-practice tests and research evaluation. 
The database structure is depicted in Figure \ref{fig:subset}.
The database schema encapsulates the relationships within a financial domain, embodying the interplay of clients, beneficiaries, and transactions. 
Organized into three distinct tables—'Source', 'Beneficiary', and 'Transactions'—the schema orchestrates the nuanced connections between primary keys, foreign keys, and regular columns. This schema not only delineates the core attributes associated with each entity but also employs color-coded annotations to differentiate primary keys, foreign keys, and standard columns.

\textit{Questions in natural language and Difficulty Level:}
We asked for expert help and manually rephrased all questions to follow ethics standards and avoid business conflicts. We created 30 questions with various difficulty levels: 8 Level-1 questions, 6 Level-2 questions, 5 Level-3 questions, and 7 Level-4 questions. 

We release this new financial dataset to the public at the following address:
\textbf{\url{https://github.com/Etamin/FinChallenge}}

\vspace{-1mm}

\usetikzlibrary{decorations.markings}

\newcommand{\PK}[1]{\cellcolor{red!30}{#1}}
\newcommand{\FK}[1]{\cellcolor{green!30}{#1}}
\newcommand{\Normal}[1]{\cellcolor{yellow!30}{#1}}
\begin{figure}[htbp]
\footnotesize
\begin{center}
\begin{tabular}{|c|}
    \hline
    \PK{Source} \\
    \hline
    \FK{Client\_ID} \\
    \Normal{Type} \\
    \Normal{Branch\_ID} \\
    \Normal{Contract\_ID} \\
    \Normal{BIC\_Code}\\
    \Normal{IBAN} \\
    \hline
\end{tabular}
\quad
\begin{tabular}{|c|}
    \hline
    \PK{Beneficiary} \\
    \hline
    \FK{Beneficiary\_ID} \\
    \Normal{Bank\_Branch\_ID} \\
    \Normal{Country\_Code} \\
    \Normal{Country\_Name} \\
    \Normal{BIC\_Code}\\
    \Normal{IBAN} \\
    \hline
\end{tabular}
\quad
\begin{tabular}{|c|}
    \hline
    \PK{Transactions} \\
    \hline
    \Normal{Transaction\_ID} \\
    \Normal{Time} \\
    \FK{Client\_ID} \\
    \FK{Beneficiary\_ID}  \\
    \Normal{Currency} \\
    \Normal{Amount} \\
    \Normal{Transaction\_Type} \\
    \hline
\end{tabular}

\end{center}
\vspace{-1mm}
\caption{Financial Dataset for Text-to-SQL (green cells indicate foreign keys)}
\label{fig:subset}
\end{figure}

\vspace{-1mm}


\subsection{Rephrased Dataset}
\label{subsec:rephrasing}
During our preliminary evaluation encompassing all SPIDER Dev datasets, it became evident that \textbf{9.4\%} of questions posed significant challenges for all models we tested, resulting in incorrect responses across diverse categories. These questions, which appear to be inadequately formed or ambiguous, can be classified as 'bad expressions'. To address this issue, two potential avenues emerge: manual rectification or leveraging automated models to curate improved question formulations; Given the tedious task of rectifying a substantial number of questions, we opted to explore the efficacy of the automated approach. For all questions of the SPIDER dataset, we choose 5 rephrased questions via ChatGPT, and if any one of them can get a better execution match and semantic similarity, we will choose this "correct" question to replace the original one.

\subsection{Experiment Settings}

\subsubsection{Environment}
We have limited computing resources, Only one NVIDIA\textsuperscript{\textregistered} DGX-1(4*V100(32GB)) can be used for inference. It includes a Intel\textsuperscript{\textregistered} Xeon\textsuperscript{\textregistered} E5-2698 v4 @ 2.20GHz CPU and 4 NVIDIA Tesla V100 GPU with 32Gib Memory.

This environment limits us to using smaller models such as the LLaMA model with 13 billion parameters for the testing. However, being restricted to the same number of parameters allows us to compare all of these models horizontally. And use int-8 inference mode if possible, which reduces GPU memory occupation to 1/4.
\subsubsection{Prompt for LLMs}
\label{sec:prompt}
In this study, we address the prompt influence challenge by designing an instructive prompt format tailored to the Text2SQL task. Our proposed format entails a structured arrangement that begins with a task description, followed by pertinent database information, and concluding with the natural language question. This format aims to guide the LLMs explicitly, providing a clear outline of the required information and context. By segmenting the prompt into distinct sections, we intend to enhance the model's understanding of the task and streamline its generation of accurate SQL queries.

In a notable case study showcasing the significant influence of prompts on Text2SQL model performance, we examined the original ALPACA model. We introduced two distinct prompt types for evaluation: Type I in Figure \ref{fig:type1}, aligned with the input format design of the T5 SOTA model, and Type II in Figure \ref{fig:type2}, structured according to the CodeX-Davinci input format. Strikingly, this change in input format led to a remarkable performance boost, with execution match rates soaring from 11.2\% to 20.8\%. This underscores the pivotal role that prompt design plays in shaping model outcomes. However, it is important to note that not all models respond equally to such prompts. Models like OpenAssistant, for instance, encounter difficulties in comprehending Type II prompts, resulting in a meager 5\% execution match rate—a clear illustration of the nuances and challenges posed by varying prompt structures in the Text2SQL task.

\begin{figure}[htpb]
\footnotesize
\begin{center}
    \fcolorbox{black}{gray!10}{\parbox{.9\linewidth}{Given the database structure as "| table\_1: column\_1, column\_2, ... | ..." , given the following database: \\
    concert\_singer | stadium : stadium\_id, location, name, capacity, highest, lowest, average | singer : singer\_id, name, country, song\_name, song\_release\_year, age, is\_male | concert : concert\_id, concert\_name, theme, stadium\_id, year | singer\_in\_concert : concert\_id, singer\_id| \\
    Give me only the SQLite query of the  question(only raw text of the code):
    }}
\end{center}
\caption{Type I}
\label{fig:type1}
\end{figure}

\begin{figure}[htpb]
\footnotesize
\begin{center}
    \fcolorbox{black}{gray!10}{\parbox{.9\linewidth}{\#\#\# \\ SQLite SQL tables, with their properties: \\
\#\\
\# stadium(Stadium\_ID, Location, Name, Capacity, ...) \\
\# singer(Singer\_ID, Name, Country, Song\_Name, ...) \\
\# concert(concert\_ID, concert\_Name, Theme, ...)\\ 
\# singer in concert(concert\_ID, Singer\_ID) \\
\# \\
\#\#\# Show name, country, age for all singers ordered by age from the oldest to the youngest.
\\
 Answer:\\
 SELECT }}
\end{center}
\caption{Type II}
\label{fig:type2}
\end{figure}

\subsubsection{Post-Processor}
Our post-processing procedure is a multi-step approach aimed at enhancing the reliability and security of the output generated by LLMs during Text2SQL tasks. In the first part, we implement a robust strategy by running the LLM five times to mitigate potential instability in its output, ensuring a more consistent and dependable result. In the second part, we meticulously filter out any extraneous elements that are not integral to the SQL query, streamlining the output for improved clarity and accuracy. Finally, in the third part, we rigorously examine the output to eliminate any potentially harmful or risky code, such as "DROP TABLE" commands, safeguarding the integrity and safety of the generated SQL queries. This comprehensive post-processing workflow not only enhances the quality of the LLM-generated SQL queries but also bolsters their security, making them more suitable for practical applications in database management and query processing.


\section{Evaluation Metrics} \label{subsec:metrics}

Let us consider the SQL query generated by the LLM under test and the ground-truth query from the dataset: we set the prediction query as token sequence $Y'$, and the ground-truth query as token sequence $Y$. The model is $f(Q)=Y'$, in which Q represents the question written in natural language. Based on Qin's survey paper~\cite{qin2022survey} on Text2SQL, we selected the evaluation metrics below.
\begin{itemize}
    \item \textbf{Exact Match}\\ Being the simplest metric in our study, Exact Match(EM) is set to 1 if $Y=Y'$, otherwise it is set to 0.
    \item \textbf{Executable} \\  Assuming the SQL execution progress on input $Y'$ is $P(Y')=Z$, and when $Z\neq NULL \| Error$, the executable rate is set to 1, otherwise 0.
    \item \textbf{Execution Match} \\ It is the most common metric when we talk about Text2SQL tasks. We set $P()$ is the execution procedure. When $P(Y)=P(Y')$ execution match is set to 1, otherwise 0.
    \item \textbf{BLEU Score} \\  The BLEU (Bilingual Evaluation Understudy) score we mentioned in the Related Work section, is a widely used metric for evaluating the quality of machine-generated translations in natural language processing and machine translation tasks. We normalize it to 0 to 1, 1 is best, 0 is worst.
\end{itemize}

The financial industry operates within a stringent regulatory framework, given the potentially devastating socio-economic repercussions that may arise from mishandling data or making risky decisions. For example, beyond financial regulations, the General Data Protection Regulation (GDPR) has a significant impact on financial institutions operating within the European Union (EU) or processing the personal data of EU residents. Therefore, SQL queries generated automatically cannot be executed on the databases without proper GDPR checks. This means that performance of Text2SQL methods should be evaluated based on static (no-execution) metrics. Such metrics will further better assess the performance of Text2SQL in a non-binary manner: some models may generate incomplete or non-executable SQL queries that are still valuable as they are very close to the ground truth, hence to what the operator wished to have. Measuring the similarity of queries, in a reliable way, appears as a promising alternative for assessment. In this work, we propose the following two metrics:
\begin{itemize}
    \item \textbf{SQAM (SQL Query Analysis Metric)} \\ This is the first SQL evaluation metric we propose, it evaluates input queries by breaking them down into their core components using regular expressions, including SELECT, FROM, WHERE, GROUP BY, HAVING, and ORDER BY clauses. It then further dissects these components into subcomponents, such as select columns and table names. The resulting accuracy score ranging from 0 to 1 is based on the subcomponents match rate.
    \item \textbf{TSED (Tree Similarity of Editing Distance)} \\ This is the second SQL evaluation metric we propose, using abstract syntax tree (AST) as features, calculating the editing difference between 2 ASTs from the prediction query and the ground-truth query. Then we calculate the TSED based on the editing distance $D$ and the node numbers $N$ of the larger AST from predict or ground truth. $TSED=D/N$. Similarly to SQAM, this metric also ranges from 0 to 1, 1 being the best and 0 the worst.
        
\end{itemize}
In the following subsections, we will introduce them in detail.
\subsection{SQAM}
The SQL Query Analysis Metric (SQAM)\footnote{https://github.com/ezzini/SQAM} evaluation method operates through a systematic process of dissecting input queries into their fundamental components, such as the SELECT, FROM, WHERE, GROUP BY, HAVING, and ORDER BY clauses, employing regular expressions. These primary segments are further deconstructed into their subcomponents, including \emph{select columns}, \emph{table names}, and \emph{where conditions}, through the application of additional regular expressions.

Subsequently, the SQAM evaluation metric computes an accuracy score by juxtaposing the subcomponents present in the input query with those found in the ground truth query. Notably, this comparison considers the hierarchical importance of each subcomponent within the query structure. For instance, components like the SELECT clause carry greater weight in the evaluation compared to others like the ORDER BY clause. The accuracy score, a crucial output of this process, is quantified as the percentage of matched subcomponents, thoughtfully weighted by their respective significance in shaping the overall query. This nuanced approach to evaluating query similarity through SQAM underscores its precision in assessing the alignment of essential query components, thereby providing a robust measure of query correspondence.

\subsection{Tree Similarity of Editing Distance (TSED)}
\begin{figure}[htb]
\centering
\includegraphics[width=\columnwidth]{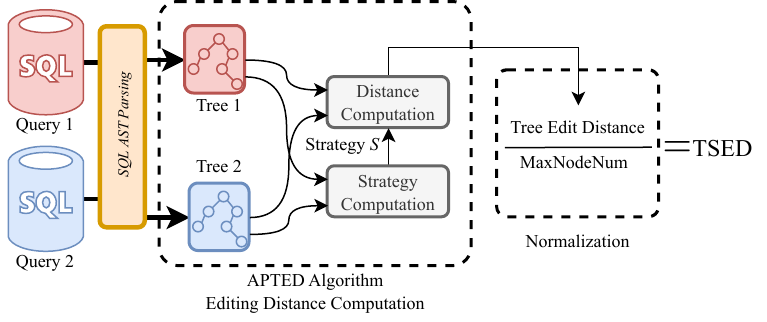}
\caption{Pipeline of TSED Evaluation Metric}
\label{fig:tsed}
\end{figure}

We propose a novel metric that we name Tree Similarity of Editing Distance (TSED)\footnote{https://github.com/Etamin/TSED} to assess the similarity between two SQL queries. This innovative approach, illustrated in Figure \ref{fig:tsed}, encompasses three distinct stages to provide a comprehensive evaluation. The initial stage involves the transformation of the two SQL queries into Abstract Syntax Trees (ASTs), facilitating a structured comparison. Following this, we calculate the editing distance between these ASTs, capturing the extent of their dissimilarity. The final step involves normalizing this distance measure to a scale between 0 and 1, allowing for a straightforward and interpretable assessment of query similarity.

\noindent 
\textbf{SQL Parsing: A Mathematical Perspective}
SQL (Structured Query Language), as a domain-specific language tailored for relational database management, inherently possesses a rich syntactic structure. The critical step of converting raw SQL text into its associated Abstract Syntax Tree (AST) is often referred to as SQL parsing. This conversion can be aptly modelled as a function, denoted as $P$. Given the space of all SQL queries,  $\mathcal{Q}$, and the corresponding space of ASTs,  $\mathcal{A}$, the function $P$ establishes a mapping between these two spaces. For any particular query $q$ that belongs to $\mathcal{Q}$, its corresponding AST is:

\begin{equation}
A(q) = P(q) \quad \text{where} \quad A(q) \in \mathcal{A} \, \text{and} \, q \in \mathcal{Q}
\end{equation}

Given the vast spectrum of SQL syntax, enriched with its numerous variations and intricacies, the process of parsing is undeniably intricate. A parser is tasked with navigating and interpreting diverse clauses, conditions, and embedded ambiguities within the SQL text.

\noindent
\textbf{Modern Solutions:} One of the contemporary tools addressing this challenge is the \texttt{@florajs/sql-parser}\footnote{https://github.com/florajs/sql-parser}. Beyond traditional parsing, this tool embeds a nuanced transformation function . If we consider \( \mathcal{S} \) as the space encompassing all SQL strings, this transformation function creates a bridge from any SQL string \( s \) in \( \mathcal{S} \) to its corresponding AST, \( a \), in \( \mathcal{A} \):

\begin{equation}
a = T(s) \quad \text{where} \quad s \in \mathcal{S} \, \text{and} \, a \in \mathcal{A}
\end{equation}
We describe its cardinality,  $|\mathcal{A}|$, using an indicator function  $\delta$ to highlight the uniqueness of each AST:

\begin{equation}
|\mathcal{A}| = \sum_{i=1}^{n} \delta(A_i) \quad \text{where} \quad \delta(A_i) = 
\begin{cases} 
1, & \text{if } A_i \text{ is unique within } \mathcal{A}\\
0, & \text{otherwise}
\end{cases}
\end{equation}


\noindent 
\textbf{Hierarchical Transformation from JSON to AST.}
Upon translating an SQL query into its AST, which is then expressed as a JSON structure, we proceed to represent this AST as a tree. The tree offers a clear hierarchical perspective, encapsulating the relational and logical intricacies of SQL queries.

We introduce a bijective function $C: J \rightarrow T$ , which is responsible for mapping every JSON object $J$ into a tree representation, $T$, where $T$ is represented as: $T = \{N, E\}$, where $N = \{n_1, n_2, ..., n_k\}$ and $E = \{e_{12}, e_{13}, ..., e_{(k-1)k}\}$.

Given this, the relationship between nodes \( n_i \) and \( n_j \) through the edge \( e_{ij} \) can be described by a matrix representation \( M \):

\begin{equation}
M(n_i, n_j) = 
\begin{cases} 
1, & \text{if } e_{ij} \in E \\
0, & \text{otherwise}
\end{cases}
\end{equation}

The tree's depth, \( depth(T) \), corresponds to the maximum nested level, \( d \), in \( J \). This depth is intrinsically linked to the longest path \( P \) in \( T \):

\begin{equation}
d(J) = \max_{\forall paths \ P} \text{length}(P)
\end{equation}

Our synthesis approach employs a recursive methodology. As we iterate over each key-value pair in \( J \), nodes are birthed in \( N \). When encountering a nested JSON object or array, the function delves recursively to ensure that the hierarchies in \( J \) and \( T \) are synchronized. Now, let \( B(n) \) represent the set of children nodes for node \( n \). The tree's branching factor, \( b \), can be formulated as: $b = \max_{n \in N} |B(n)|$.


\begin{table*}[htbp]
\resizebox{.75\linewidth}{!}{%
\begin{tabular}{l|lrrrrrr}
\hline
Model Type                         & Model Name      & Parameter Size         & Level 1                   & Level 2                   & Level 3 & Level 4 & All   \\ \hline
\multirow{10}{*}{General LLM}      & ChatGPT-3.5-turbo    & 175B    & 0.760                     & 0.799                     & 0.408   & 0.493   & 0.623 \\
                                   & DIN-SQL+GPT-4     & 1.76T       & 0.861                     & 0.866                     & 0.700   & 0.654   & \textbf{0.762} \\
                                   & CodeX-Davinci-3    & 175B      & 0.730                     & 0.799                     & 0.392   & 0.382   & 0.570 \\ \cline{2-8} 
                                   & MPT-7B-instruct     &  7B   & 0.262                     & 0.381                     & 0.117   & 0.091   & 0.205 \\
                                   & ALPACA         &   7B    & 0.311                     & 0.460                     & 0.192   & 0.083   & \textbf{0.242} \\
                                   & KOALA     &      7B      & 0.195                     & 0.218                     & 0.017   & 0.071   & 0.131 \\
                                   & OpenAssistant-pythia & 12B & 0.202                     & 0.322                     & 0.025   & 0.069   & 0.157 \\
                                   & ORCA-mini     &   7B     & 0.243                     & 0.280                     & 0.101   & 0.076   & 0.169 \\
                                   & LLaMA-2     &    7B      & 0.225 & 0.393 & 0.101   & 0.081   & 0.192 \\ \hline
\multirow{4}{*}{Code Specific LLM} & CodeGen2      &    7B    & 0.375                     & 0.498                     & 0.167   & 0.066   & 0.257 \\
                                   & Starcoder   &   15.5B    & 0.584                     & 0.628                     & 0.275   & 0.208   & 0.410 \\
                                   & Vicuna      &    7B    & 0.060                     & 0.134                     & 0.008   & 0.042   & 0.064 \\
                                   & nsql       &    6B       & 0.772                     & 0.732                     & 0.608   & 0.277   & \textbf{0.548} \\ \hline
\multirow{3}{*}{Seq-to-Seq Model}  & T5(tscholak/cxmefzzi) & 3B & 0.828                     & 0.782                     & 0.650   & 0.434   & 0.641 \\
                                   & PICARD+T5    &    3B     & 0.790                     & 0.799                     & 0.558   & 0.502   & 0.652 \\
                                   & RESDSQL          &   3B     & 0.872                     & 0.857                     & 0.666   & 0.696   & \textbf{0.775} \\ \hline
\end{tabular}
}
\caption{Benchmark Results of Execution Match of all Models we tested on the "dev" SPIDER dataset}
\label{table:execution}
\vspace{-4mm}
\end{table*}

\noindent
\textbf{Tree Distance Computation.}
One of the fundamental challenges in comparing trees, especially those derived from SQL queries, is quantifying their similarity or difference. Here, the concept of tree edit distance comes into play.

The tree edit distance between two trees \( T_1 \) and \( T_2 \) measures the minimum number of basic edit operations (insertion, deletion, and renaming of nodes) required to transform \( T_1 \) into \( T_2 \). Mathematically, the tree edit distance function \( \Delta \) can be defined as:
\vspace{-0.1cm}
\begin{equation}
\Delta(T_1, T_2) = \min_{ops} \sum_{i=1}^{n} w(op_i)
\end{equation}
where  \textit{ops} is a sequence of \textit{n} edit operations that transform  $T_1$ to $T_2$, and $w(op_i)$ is the cost associated with the  $i^{th}$ operation. The goal is to find the sequence of operations that minimizes the total cost.

However, computing the tree editing distance is computationally intensive. The problem is NP-hard, and naive algorithms have a time complexity that is doubly exponential in the size of the trees. Let $|T_1|$ and $|T_2|$ denote the sizes of trees $T_1$ and $T_2$ respectively. The computational complexity $\mathcal{C}$ can be represented as:

\begin{equation}
\mathcal{C}(|T_1|, |T_2|) = \mathcal{O}\left(2^{2^{|T_1| + |T_2|}}\right)
\end{equation}

Given the need for efficient computation, we employ the `APTED' algorithm~\cite{pawlik2016tree}~\cite{pawlik2015efficient}, a state-of-the-art tool designed for tree edit distance computation. The library optimizes the process using advanced algorithms and data structures, significantly reducing computational time. In our project, the `APTED' function compute the distance, the return $\delta$ will be an integer:
$\delta = APTED(T_1, T_2)$, where  $\delta$ represents the tree edit distance between $T_1$ and $T_2$.

In essence, tree edit distance provides a robust metric to quantify the similarity between SQL query structures, and the `apted' library facilitates its efficient computation, enabling a deeper understanding and comparison of different SQL queries.

\textbf{Normalization.}

Beyond merely calculating the raw editing distance, we take into account the complexity of the queries by normalizing the distance using the maximum number of tree nodes:  
\begin{equation}
TSED = \delta/ MaxNodes(T_1, T_2)
\end{equation}
This normalization allows us to obtain a more comprehensive and meaningful final result, as it accounts for variations in query complexity and structure. By computing the distance-to-maximum-nodes ratio, we arrive at a final evaluation metric that offers a nuanced perspective on the accuracy and efficiency of our Text2SQL conversion techniques. This approach ensures that our assessments are not only sensitive to the accuracy of the conversions but also consider the relative complexity of queries, enabling a more balanced and informative measure of performance.

\section{Experiment Results}
\label{sec:exp}
\subsection{Benchmarks}

\textbf{Execution Match Score on the SPIDER dataset}

In our experimentation, we organized the models into three distinct groups as illustrated in Table \ref{table:execution}: general purpose LLMs, Code-Specific LLMs, and Sequence-to-Sequence models. 
Table \ref{table:execution} further presents the Execution Match score on the SPIDER dataset for each studied LLM and for each of the four difficulty levels. 
Note that for all LLMs, we run our experiments with both Type I and Type II prompts (cf. \ref{sec:prompt}), and we always select best performance.
The overall winner is the GPT-4 + DIN approach which emerged as the most effective choice across all General LLMs. 
Furthermore, when focusing on models with fewer than 7 billion parameters, ALPACA stood out as the top-performing option following prompt optimization. 

Within the Code-Specific LLMs group, nsql-6b took the lead as the superior model, with the entire subgroup showcasing significantly improved performance compared to other LLM models of a similar parameter size. This finding underscores the value of specialized models in handling domain-specific tasks effectively.

Meanwhile, the Seq-to-Seq models demonstrated a consistently high level of performance, closely resembling the capabilities of models like RESDSQL. However, we anticipate that these Seq-to-Seq models may face formidable competition in the near future.

\begin{tcolorbox}[leftrule=0mm,rightrule=0mm,toprule=0mm,bottomrule=0mm,left=0pt,right=0pt,top=0pt,bottom=0pt,title={RQ1: Which LLM is the most accurate for the Text2SQL task?}]
  \textbf{Answer: } The super giant GPT-4 model is the best-performing LLM. However, the results suggest that specific training datasets and "focused" LLMs can also perform well. 
  For example, nsql-6b performs better than other models with 7 billion parameters. This means that general LLM is not a silver bullet. 
  We also notice that the Level-4 questions are always hard for every model, and Level-1 and Level-2 questions usually lead to similar performances for these models.

\end{tcolorbox}

\textbf{Time and Memory performances of Three typical LLMs}

To investigate the time and memory performance of LLMs, we selected three typical LLMs that we can download and run on our local machine. We selected T5 (with 3B parameters) and two versions of LLaMA (7B and 70B parameters).

As indicated in Table~\ref{table:time} above, it is evident that the 70-billion-parameter model consumes a substantial 70 gigabytes of GPU memory, rendering it impractical for use on our modest DGX workstation with limited memory capacity. Furthermore, the inference time for this model exceeds a staggering 100 seconds per instance. These findings underscore the critical importance of evaluating the performance and resource requirements of LLMs, as they demand considerable computational resources that may not be feasible or efficient for certain hardware setups and use cases.

\begin{table}[H]
\resizebox{.65\linewidth}{!}{%
\begin{tabular}{l|ll}
\hline
Model & Time       & GPU Memory  \\ \hline
T5-3B(fp32)   & $\sim$1.6s & $\sim$12GiB \\ \hline
LLaMA-7B(int-8)  & $\sim$5.5s & $\sim$7GiB  \\ \hline
LLaMA-70B(int-8) & $\sim$100s & $\sim$70GiB \\ \hline
\end{tabular}
}
\caption{Inference Time \& Memory for Different Parameters}
\label{table:time}
\end{table}

\begin{tcolorbox}[leftrule=0mm,rightrule=0mm,toprule=0mm,bottomrule=0mm,left=0pt,right=0pt,top=0pt,bottom=0pt,title={RQ3: Which LLM is the most efficient regarding execution time and computational resources?}]
  \textbf{Answer: } We highly recommend the adoption of the 7-billion-parameter model for small-sized businesses or research groups. When considering the performance of nsql-6b, with dedicated attention to training data and prompt optimization, the 6-billion-parameter model can surpass the capabilities of certain 13-billion-parameter or larger models. This presents a cost-effective and efficient alternative for organizations seeking high-performing language models without the extensive computational overhead associated with larger counterparts.
\end{tcolorbox}


\subsection{Rephrase SPIDER}


This experiment aims to assess whether rephrasing a question helps generate better SQL queries. 
As explained in Section \ref{subsec:rephrasing}, thanks to ChatGPT, we first generate 5 variants of the initial questions. 
One critical challenge to conduct this experiment is that a typical LLM does not generate always the same answer, even if the question is always the same. 
Consequently, if a generated SQL query is better with a rephrased question, we do not know whether the query is better because of the rephrasing or because of the random nature of the LLM. 
To overcome this issue, we perform this experiment with the only model that stays "constant", i.e., which always generates the same answer each time we ask the same question. This model is PICARD. 
According to the result presented in Table \ref{table:rephrase}, the dataset with rephrased questions yields better results than the original SPIDER dataset. 
This result suggests that a research direction on question optimization is worthy of consideration. Oftentimes, models just respond "badly" to "bad" problems.

\begin{table}[H]
\begin{tabular}{l|rrrr}
\hline
Batch    & \multicolumn{1}{l}{Exec Match} & \multicolumn{1}{l}{BLEU} & \multicolumn{1}{l}{SQAM} & \multicolumn{1}{l}{TSED} \\ \hline
Original & 0.651                          & 0.249                    & 0.847                    & 0.617                    \\ \hline
Rephrase & \textbf{0.797}                          & \textbf{0.271}                    & \textbf{0.891}                    & \textbf{0.725}                    \\ \hline
\end{tabular}
\caption{Rephrase Results on PICARD}
\label{table:rephrase}
\end{table}
\vspace{-0.5cm}

\begin{tcolorbox}[leftrule=0mm,rightrule=0mm,toprule=0mm,bottomrule=0mm,left=0pt,right=0pt,top=0pt,bottom=0pt,title={RQ2: What is the effect of the question rephrasing, prompt optimization, and post-processing steps on the performance of the LLM?}]
  \textbf{Answer: } Results from Sections 4.5 and this SubSection suggest that optimizing questions and prompts is very effective and can significantly improve performance in certain scenarios. Performance improvements of 5\% and 14\% were achieved on ALPACA and PICARD respectively. 
\end{tcolorbox}

\subsection{TSED \& SQAM}
In this experiment, we study the effectiveness of the two metrics we proposed, TSED and SQAM, in measuring the quality of the generated SQL queries.
We consider a metric as effective if the metric is "close" to the Execution Match metric, which is the best metric (among the ones we consider in this study) to measure semantic similarity between a generated SQL query and the ground truth SQL query.
As a reminder, as explained in Section \ref{subsec:metrics}, the execution match metric cannot always be used under the constraints of our industrial partner. This is why we proposed TSED and SQAM, which can be computed without executing the generated queries. 

\begin{figure}[htbp]
\centering
\includegraphics[width=\columnwidth]{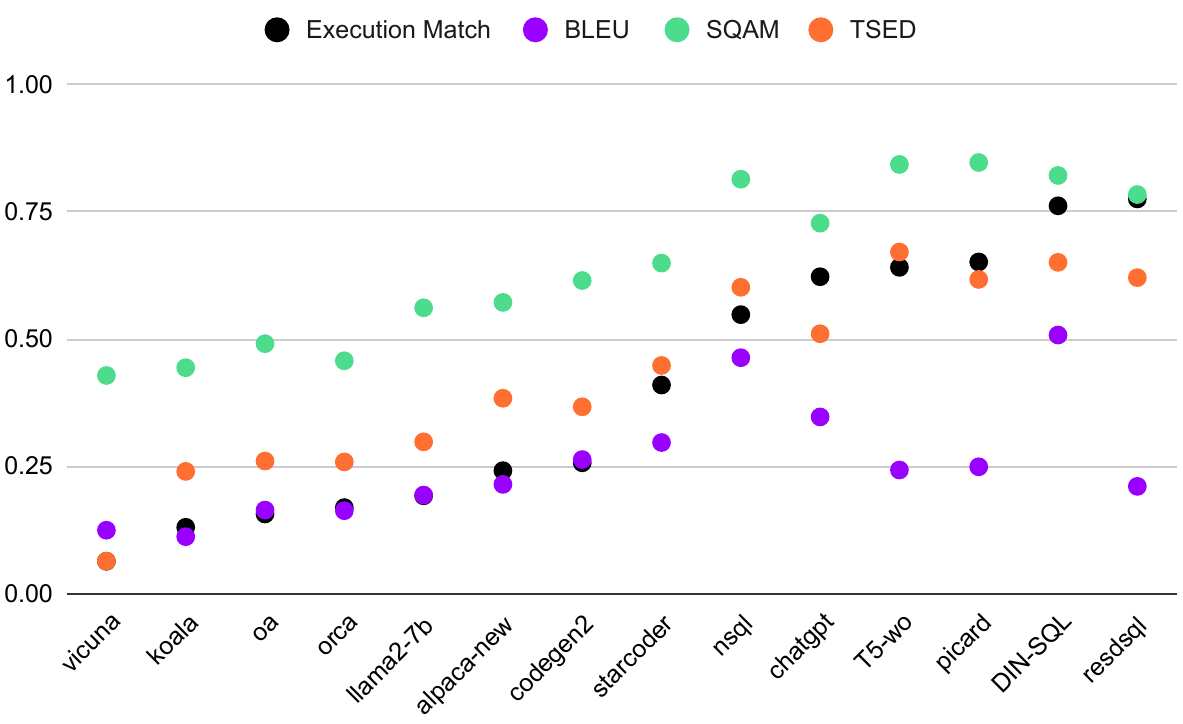}
\caption{Distribution of Evaluation Metrics}
\label{fig:distribution}
\end{figure}

TSED and SQAM scores (as well as Bleu Score, Exact Match, and Executable) 
are presented in Table \ref{table:metrics_all}. 
Impressively, both TSED and SQAM metrics exhibited very high Pearson product-moment correlation coefficients (PCCs) with execution match, averaging around an exceptional 0.95  as shown in Table \ref{table:corref}. Notably, TSED demonstrated a slight edge over SQAM in terms of PCC. However, what sets TSED apart is its remarkable performance in the SSD (Sum of Squares of Deviations) metric, where lower values are indicative of superior accuracy. As depicted in Figure \ref{fig:distribution}, both TSED and SQAM exhibited trends closely aligned with execution match, with TSED displaying an even closer correlation. Notably, we observed that the BLEU score for the higher performance part proved to be unreliable in this context.

\begin{table}[htbp]
\begin{tabular}{lrrrrr}
\hline
Model Name        & \multicolumn{1}{l}{BLEU}  & \multicolumn{1}{l}{SQAM} & \multicolumn{1}{l}{TSED} & EAble & EM \\ \hline
ChatGPT-3.5 & 0.347                     & 0.728                    & 0.510                    & 0.885       & 0.219       \\
DIN-SQL+GPT4           & \textbf{0.508}                     & \textbf{0.821}                    & \textbf{0.651}                    & \textbf{0.988}       & \textbf{0.319}       \\
Davinci           & 0.348                     &  0.564           &  0.575             & 0.844       & 0.235       \\  \hline
MPT-7B            & 0.193                     & 0.557                    & 0.343                    & 0.465       & 0.020       \\
ALPACA-7B         & \textbf{0.205}                     & \textbf{0.572}                    & \textbf{0.384}                    & \textbf{0.552}       & \textbf{0.126}      \\
KOALA-7B          & 0.109                     & 0.444                    & 0.240                    & 0.305       & 0.022       \\
OpenAssistant     & 0.160                     & 0.491                    & 0.260                    & 0.348       & 0.050       \\
ORCA              & \multicolumn{1}{l}{0.163} & 0.457                    & 0.258                    & 0.399       & 0.047       \\
LLaMA-2           & \multicolumn{1}{l}{0.193} & 0.561                    & 0.298                    & 0.417       & 0.074       \\ \hline
CodeGen2-7B       & 0.264                     & 0.615                    & 0.367                    & 0.583       & 0.124       \\
Starcoder-15.5B   & 0.287                     & 0.649                    & 0.448                    & \textbf{0.769}       & 0.160       \\
Vicuna-7B         & 0.122                     & 0.428                    & 0.063                    & 0.102       & 0.026       \\
nsql-6B           & \textbf{0.443}                     & \textbf{0.814}                    & \textbf{0.602}                    & 0.741       & \textbf{0.308}       \\ \hline
T5-3B             & 0.243                     & 0.843                    & \textbf{0.671}                    & 0.863       & \textbf{0.409}       \\
PICARD+T5/3B      & \textbf{0.249}                     & \textbf{0.847}                    & 0.617                    & 0.904       & 0.371       \\
RESDSQL           & 0.210                     & 0.784                    & 0.621                    & \textbf{0.998}       & 0.037       \\ \hline
\end{tabular}
\caption{Evaluation Metrics: Bleu Score, SQAM, TSED, Executable (EAble), and Exact Match (EM), on the SPIDER dataset. Scores are computed as an average of the scores for the 4 question difficulty levels.}
\label{table:metrics_all}
\end{table}

In a noteworthy special case, we observe that TSED (Tree Similarity for Editing Distance) proves to be more effective than BLEU when evaluating the similarity of SQL queries. Let us consider two SQL queries: Query 1: 

{\footnotesize
\begin{code}
SELECT stadium.name, count() 
FROM concert 
JOIN stadium 
ON concert.Stadium_ID = stadium.Stadium_ID 
GROUP BY concert.stadium_id;
\end{code}
}

and Query 2:
{\footnotesize
\begin{code}
SELECT T2.name, count() 
FROM concert AS T1 
JOIN stadium AS T2 
ON T1.stadium_id = T2.stadium_id 
GROUP BY T1.stadium_id;
\end{code}
}

\begin{table}[h]
\begin{tabular}{l|lll}
\hline
    & BLEU & SQAM  & TSED  \\ \hline
$PCC\uparrow $ & 0.681 & 0.948 & \textbf{0.951} \\ \hline
$SSD\downarrow $ & 0.804 & 1.022 & \textbf{0.129} \\ \hline
\end{tabular}
\caption{Correlation and statistical difference between 3 evaluation metrics and Execution Match}
\label{table:corref}
\end{table}

Remarkably, both queries achieve a perfect execution match with a score of 1.0, indicating that they produce identical results when executed. However, when assessed using the TSED metric, these queries exhibit a similarity score of 0.864, signifying a high degree of structural similarity in their abstract syntax trees (ASTs). In contrast, the BLEU score, which ranges from 0 to 1, reaches only a modest 0.257. This intriguing case underscores the nuanced nature of query evaluation metrics and highlights how TSED excels at capturing structural similarities that BLEU may overlook.

\begin{tcolorbox}[leftrule=0mm,rightrule=0mm,toprule=0mm,bottomrule=0mm,left=0pt,right=0pt,top=0pt,bottom=0pt,title={RQ4: When the database is not reachable, can we propose metrics to evaluate the quality of the generated SQL queries that are as accurate as the Execution Match metric?}]
  \textbf{Answer: } TSED is a good metric that can effectively evaluate SQL queries without access to the database.
\end{tcolorbox}


\subsection{Experiments on the New Financial Dataset}

For this experiment, we selected ChatGPT(GPT-4), ALPACA, and nsql-6B, which performed the best according to Table \ref{table:execution} for their respective Model Type group. 
Regarding Seq-to-Seq models, we selected PICARD+T5 instead of RESDSQL because RESDSQL performed poorly on the new financial dataset.  

We run the 4 LLMs on our financial dataset for the Text2SQL task, and Table \ref{table:newdataset} presents the results. 
Despite some models being specially designed for SQL generation in natural language processing, ChatGPT continues to stand as the state-of-the-art (SOTA) model for this task. However, when confronted with the challenges posed by this distinct dataset, which does not conform to the structured format of the SPIDER dataset, our model comparison revealed some intriguing results. Specifically, the PICARD model, which has shown exceptional promise in previous evaluations, failed to reach the same level of performance as ChatGPT. Even more notably, our minor dataset proved to be a formidable challenge for both language models, including non-Language Models (non-LLMs) and Language Models (LLMs). None of the models achieved an execution match surpassing the 70\% threshold, underscoring the complexity of our bank transaction sub-dataset and the need for further research and innovation in Text2SQL tasks under non-standardized domains.
\begin{table}[htbp]
\begin{tabular}{l|llll}
\hline
Model       & BLEU  & SQAM  & TSED  & EM    \\ \hline
ChatGPT-3.5 & \textbf{0.582} & 0.696 & \textbf{0.599} & \textbf{0.633} \\
ALPACA-7B   & 0.417 & 0.596 & 0.425 & 0.300 \\
NSQL-6B     & 0.444 & 0.678 & 0.486 & 0.433 \\
PICARD-3B   & 0.057 & \textbf{0.705} & 0.484 & 0.566 \\ \hline
\end{tabular}
\caption{Results of Banking sub-dataset for Text2SQL, EM is Execution Match}
\label{table:newdataset}
\end{table}

\begin{tcolorbox}[leftrule=0mm,rightrule=0mm,toprule=0mm,bottomrule=0mm,left=0pt,right=0pt,top=0pt,bottom=0pt,title=RQ5: Text2SQL tasks in financial scenarios (especially banks) are lacking in existing datasets. How should we solve it?]
  \textbf{Answer: }  We introduce an efficient new financial dataset and assess it using several state-of-the-art (SOTA) models. This newly crafted dataset poses a substantial challenge, featuring a notably higher incidence of level 4 complexity problems compared to the widely recognized SPIDER dataset. This observation underscores the significance of encouraging companies to engage in the creation and evaluation of such sub-datasets.
\end{tcolorbox}

\section{Threats to Validity} \label{sec:threats}
The validity concerns most pertinent to our evaluation are internal and external validity.
\sectopic{Internal Validity. }
In this research, internal validity would encompass ensuring that the differences in performance observed among the selected LLMs on the provided datasets can be confidently attributed to the variations in model capabilities and not influenced by confounding factors or unintended biases. In the realm of internal validity, the research strategically harnesses multi-time output averaging to mitigate the inherent variability in LLM outputs proposed by Tian et al ~\cite{tian2023chatgpt}. Recognizing the potential for erratic responses, this approach tempers the impact of occasional instability, thereby fostering more reliable and consistent performance measurements. Furthermore, the study introduces a meticulously designed post-processor tailored to the idiosyncrasies of LLM output. 
\sectopic{External Validity.} 
External validity would involve evaluating whether the outcomes observed among the chosen LLMs on the given datasets can be extended to a broader range of scenarios, datasets, or real-world applications. On the front of external validity, the research extends its purview to encompass banking scenarios, thereby transcending the initial domain boundaries. By juxtaposing the performance of the selected LLMs against the original SPIDER dataset, the study ventures beyond its initial context and broadens the applicability of its findings. 
\sectopic{Construct Validity. } 
Construct validity would involve ensuring that the chosen LLMs, evaluation metrics, and datasets accurately capture the dimensions of interest related to Text2SQL performance. In terms of construct validity, the research takes a comprehensive approach by evaluating the performance of LLMs across five distinct evaluation metrics. By encompassing multiple dimensions of assessment, the study robustly captures the nuances of LLM-generated SQL queries. 
\section{Conclusion} \label{sec:conclusion}

In this paper, we investigated five research questions focusing on the practice of the text2SQL models in financial scenarios. 
We evaluate a number of LLMs on the SPIDER dataset for the text2SQL task, study challenges from financial practice, evaluate metrics that do not need to run the SQL queries, and optimize some of the questions and prompts. 
We contribute a benchmark that can be used for model selection, 2 open-source evaluation metrics called SQAM and TSED, and present a small text2SQL challenge of the bank database. We find three main research findings: 
\ding{182} We found that in the case of general-purpose 7B LLMs perform poorly on Text2SQL tasks and require more SQL-related training data. The nsql-6B LLM appears to be a good alternative to giant general-purpose LLMs.  
\ding{183} The dataset needs to be optimized, especially the quality of questions, and find effective prompts. 
\ding{184} The TSED metric is effective in measuring the quality of generated SQL queries. \\
In our ongoing project, we have exciting plans for future work. We intend to extend the application of our innovative evaluation metrics beyond the banking domain and apply them to major code generation tasks, thereby broadening the scope of their utility. Additionally, we are in the process of developing an automatic toolkit tailored for Text-to-Code evaluation. This toolkit will not only streamline the evaluation process but also contribute to the standardization of evaluation procedures in the field.

\sectopic{Acknowledgement.} This work was funded by Luxembourg's National Research Fund (FNR) under the grant BRIDGES. We are grateful to the DataLab team at BGL BNP Paribas for their valuable insights and assistance.


\bibliographystyle{ACM-Reference-Format}
\balance
\bibliography{paper}

\end{document}